\documentclass[journal]{IEEEtran}
\usepackage{amsmath,amsfonts,amssymb}
\usepackage{algorithmic}
\usepackage{array}
\usepackage[caption=false,font=footnotesize]{subfig}
\usepackage{textcomp}
\usepackage{stfloats}
\usepackage{url}
\usepackage{verbatim}
\usepackage{graphicx}
\usepackage{multirow}
\usepackage{siunitx}
\DeclareSIUnit[quantity-product = {}] \baud{\text{baud}}
\usepackage[acronym,nomain]{glossaries}
\glsdisablehyper

\usepackage[shortcuts]{extdash}
\newacronym{AWGN}{AWGN}{Additive White Gaussian Noise}
\newacronym{BPSK}{BPSK}{Binary Phase-Shift Keying}
\newacronym{CAZAC}{CAZAC}{Constant Amplitude Zero Auto-Correlation}
\newacronym{DAS}{DAS}{Distributed Acoustic Sensing}
\newacronym{DSP}{DSP}{Digital Signal Processing}
\newacronym{KF}{KF}{Kalman Filter}
\newacronym{LMS}{LMS}{Least Mean Square}
\newacronym{LS}{LS}{Least Squares}
\newacronym{LFSR}{LFSR}{Linear-Feedback Shift Register}
\newacronym{LO}{LO}{Local Oscillator}
\newacronym{m-sequence}{m-sequence}{Maximum Length Sequence}
\newacronym{O-ISAC}{O-ISAC}{optical-fiber-based integrated sensing and communication}
\newacronym{OTDR}{OTDR}{Optical Time Domain Reflectometry}
\newacronym{PDF}{PDF}{Probability density function}
\newacronym{PNC}{PNC}{Phase Noise compensation}
\newacronym{PPA}{PPA}{Perfect Periodic Autocorrelation}
\newacronym{PRBS}{PRBS}{Pseudorandom Binary Sequence}
\newacronym{PS-MP}{PS-MP}{Perfect-Square Minimum-Phase}
\newacronym{PSK}{PSK}{Phase Shift Keying}
\newacronym{PSR}{PSR}{Peak to Sidelobe Ratio}
\newacronym{RBS}{RBS}{Rayleigh backscattering}
\newacronym{RMSE}{RMSE}{Root Mean Squared Error}
\newacronym{RSOP}{RSOP}{Rotation of the State of Polarization}
\newacronym{SNR}{SNR}{Signal-to-Noise Ratio}
\newacronym{SSMF}{SSMF}{Standard Single-Mode Fiber}
\newacronym{SMF}{SMF}{Single-Mode Fiber}
\newacronym{STD}{STD}{Standard Deviation}
\newacronym{ZC}{ZC}{Zadoff-Chu}
\def\BibTeX{{\rm B\kern-.05em{\sc i\kern-.025em b}\kern-.08em
    T\kern-.1667em\lower.7ex\hbox{E}\kern-.125emX}}
\usepackage{balance}
\begin{document}
\title{Comparative Analysis of Spread-Spectrum Codes for Fibre-Optic Distributed Acoustic Sensing}
\author{Maria Freire-Hermelo and Élie Awwad
\thanks{Manuscript created February, 2026.

Maria Freire-Hermelo and Élie Awwad are with the COMELEC department, LTCI, Télécom Paris, IP Paris, 19 place Marguerite Perey, Palaiseau, France (e-mail: maria.freirehermelo@telecom-paris.fr, elie.awwad@telecom-paris.fr).}}

\markboth{}
{How to Use the IEEEtran \LaTeX \ Templates}

\maketitle

\begin{abstract}
This paper presents a comprehensive performance comparison of several coded sequences for distributed acoustic sensing systems through numerical simulations. Correlation analysis reveals that perfect autocorrelation sequences provide accurate channel estimation. Phase noise sensitivity varies significantly with frequency diversity, with high-diversity sequences maintaining lower estimation errors for phase and intensity despite increased laser linewidths. For a coded distributed acoustic sensing system with \qty{10}{\hertz} laser linewidth, high-diversity sequences achieve $2.7\times10^{-3}$ rad phase error compared to $5.3\times10^{-3}$ rad for low-diversity codes for a \qty{1.5}{\kilo\meter} fibre length. The findings establish performance trade-offs between correlation quality, phase noise resilience, and practical implementation constraints for next-generation fibre-optic distributed acoustic sensing systems.
\end{abstract}


\section{Introduction}

\IEEEPARstart{R}{ecently}, \gls{DAS} has emerged as a versatile technology for fibre-optic sensing, enabling continuous infrastructure monitoring and providing timely alerts for different events, such as earthquake detection~\cite{fernandez_2020_distributed}. Compared to pulsed laser interrogation techniques, whose measurement sensitivity is limited by high peak power requirements, continuous-wave \gls{DAS} systems employing coded sequences to modulate the transmitted signal, also known as spread-spectrum \gls{DAS}, have shown promise in reducing peak power, thereby mitigating nonlinear crosstalk and enabling potential coexistence of fibre communications and sensing on the same fibre~\cite{choudhury_2025_wavelength}. Such coexistence of data transmission and \gls{DAS} over a single optical fibre exemplifies \gls{O-ISAC}, where sensing and communications share the same infrastructure~\cite{he2023integrated}.

To unlock the potential of \gls{O-ISAC}, optimal coded sequences are essential for reliable \gls{DAS} performance.
Yet determining the best choice for transmission in \gls{DAS} systems remains an open research problem. Previous studies have investigated the use of Legendre, \gls{m-sequence} and Golay codes~\cite{martins_real_2016, mompo_distributed_2019, dorize_enhancing_2018}; however, comprehensive comparative analyses across different sequence types are still limited. To date, only one study has numerically compared the performance of Golay complementary sequences with that of \gls{PS-MP} \gls{CAZAC} sequences~\cite{dorize_optimal_2020}.

Here, we present a detailed evaluation of various coded sequences applied within a \gls{DAS} system, in particular within the coherent $\Delta\phi$-OTDR implementation of DAS, focusing particularly on laser phase noise, one of the main limitations of such systems~\cite{wang2015influence, awwad2020detection}. Section~\ref{sec:selected_sequences} introduces the selected sequences and discusses their key properties. Section~\ref{sec:seq_das} analyses their performance using a simulation framework under different laser phase noise conditions. Numerical analysis confirms the impact of laser phase noise on performance for all sequences, revealing differences that depend on the frequency diversity characteristics of each sequence. For a \qty{2.55}{\kilo\meter} fibre length, more frequency-diverse sequences achieve lower phase errors of $4.5 \times 10^{-3}$~rad compared to $7.2 \times 10^{-3}$~rad for less frequency-diverse sequences.

\section{Coded sequences}\label{sec:selected_sequences}

The following coded sequences, commonly employed in previous \gls{DAS} studies, are considered in this work to compare their performance under various laser phase noise conditions~\cite{martins_real_2016, mompo_distributed_2019, dorize_enhancing_2018}. In all definitions below, $\mathrm{N}$ denotes the sequence length, whose exact value depends on the type of sequence considered:

\begin{itemize}

    \item \gls{m-sequence}: A maximum length \gls{PRBS} generated by a \gls{LFSR}, used in several \gls{DAS} studies~\cite{martins_real_2016, mompo_distributed_2019}. Encoded in the set $\{-1, +1\}$, with sequence length $\mathrm{N} = 2^\mathrm{K} - 1$ for a register length $\mathrm{K}$, achieving the maximum possible period. No padding or truncation is considered, as it would degrade autocorrelation properties.
    
    \item Legendre: Also applied in \gls{DAS} experiments~\cite{mompo_distributed_2019}, a Legendre sequence of length $\mathrm{N}$, where $\mathrm{N}$ is an odd prime of the form  $\mathrm{N} = 4r - 1$ ($r$ being an integer) is determined as~\cite{yang2006modified, rosen2011elementary}:

    \begin{equation}
    c_n =
    \begin{cases}
    \;\;\;1, & \text{if } n^{\frac{\mathrm{N}-1}{2}} \equiv 1 \pmod{\mathrm{N}}, \\[6pt]
    -1, & \text{otherwise}.
    \end{cases}
    \end{equation}
    
    \item Golay complementary sequences: Adopted in \gls{DAS} systems, these real-valued sequences can also be defined over the alphabet $\{-1, +1\}$. Pairs of mutually orthogonal Golay sequences exist and have been exploited for dual-polarisation probing~\cite{dorize_enhancing_2018}. Golay complementary sequences can be constructed recursively from smaller Golay sequences. For length $\mathrm{N} = 2^{p + 2}$ where $p \ge 1$ is the recursion index (starting from base sequences of length 4), we denote $\{\boldsymbol{G}^{(\mathrm{N})}_{a_1}, \boldsymbol{G}^{(\mathrm{N})}_{b_1}\}$ and $\{\boldsymbol{G}^{(\mathrm{N})}_{a_2}, \boldsymbol{G}^{(\mathrm{N})}_{b_2}\}$ as two such orthogonal pairs, constructed recursively as~\cite{dorize_enhancing_2018}:

    \begin{equation}
        \begin{aligned}
            &\left\{
            \begin{aligned}
            \boldsymbol{G}^{(\mathrm{N})}_{a_1} &= \bigl[\, \boldsymbol{G}^{(\mathrm{N}/2)}_{a_1},\; \;\boldsymbol{G}^{(\mathrm{N}/2)}_{b_1} \,\bigr], \\[4pt]
            \boldsymbol{G}^{(\mathrm{N})}_{b_1} &= \bigl[\, \boldsymbol{G}^{(\mathrm{N}/2)}_{a_1},\; -\boldsymbol{G}^{(\mathrm{N}/2)}_{b_1} \,\bigr], \\[4pt]
            \boldsymbol{G}^{(\mathrm{N})}_{a_2} &= \bigl[\, \boldsymbol{G}^{(\mathrm{N}/2)}_{a_2},\; \;\boldsymbol{G}^{(\mathrm{N}/2)}_{b_2} \,\bigr], \\[4pt]
            \boldsymbol{G}^{(\mathrm{N})}_{b_2} &= \bigl[\, \boldsymbol{G}^{(\mathrm{N}/2)}_{a_2},\; -\boldsymbol{G}^{(\mathrm{N}/2)}_{b_2} \,\bigr].
            \end{aligned}
            \right.
        \end{aligned}
    \end{equation}
    
    In transmission, the pair $\{\boldsymbol{G}^{(\mathrm{N})}_{a_1}, \boldsymbol{G}^{(\mathrm{N})}_{b_1}\}$ is sent sequentially over one polarization, while $\{\boldsymbol{G}^{(\mathrm{N})}_{a_2}, \boldsymbol{G}^{(\mathrm{N})}_{b_2}\}$ is sent over the orthogonal polarization, enabling polarization-multiplexed Golay probing~\cite{dorize_enhancing_2018}.

    \item \gls{PS-MP} \gls{CAZAC}: These sequences have been previously applied for channel estimation in optical communication systems and proposed for \gls{DAS} applications~\cite{dorize_enhancing_2018, pittala_efficient_2012}. For a \gls{PS-MP} \gls{CAZAC} sequence of length $\mathrm{N}$, where $\mathrm{N}=4^\mathrm{M}$ with $\mathrm{M}$ being a non-zero positive integer, each symbol $c_n$ is defined as:

    \begin{equation}
        c_n = e^{  j \frac{2 \pi}{\sqrt{\mathrm{N}}} \biggl( \mathrm{mod} \bigl(n-1, \sqrt{\mathrm{N}}\bigr)+1 \biggr) \biggl( \biggl\lfloor\frac{n-1}{\sqrt{\mathrm{N}}}\biggr\rfloor + 1 \biggr) },
    \end{equation}

    where $n$ ranges from 0 to $\mathrm{N} - 1$. The resulting complex symbols have constant amplitude and uniformly distributed phase over the unit circle.
    
    \item \gls{ZC} \gls{CAZAC}: Another type of \gls{CAZAC} sequence, the \gls{ZC} sequence provides complex symbols of constant amplitude with phases over the unit circle. Each symbol $c_n$, for $n$ ranging from 0 to $\mathrm{N} - 1$, is defined as:

    \begin{equation}\label{eq:zc}
        c_n = e^{-j \pi q \frac{n ( n + 1 ) }{\mathrm{N}} },
    \end{equation}

    where $q \in \{1, 2, \dots, \mathrm{N}-1\}$ is called the root index, generating a distinct \gls{ZC} sequence (chosen coprime with $\mathrm{N}$ for optimal correlation properties)~\cite{chu1972polyphase, andrews2022primer}. Modifying $q$ adjusts the sequence frequency diversity, i.e. how broadly the sequence energy spreads across frequencies. The following analysis uses three \gls{ZC} sequences with distinct $q$ values yielding minimum ($q=1$), medium ($q=16269$), and maximum ($q=3036$) frequency diversity, as shown in the spectrograms in Section~\ref{sec:seq_b2b_time_freq}.
    
\end{itemize}

\gls{CAZAC} sequences, including \gls{PS-MP} and \gls{ZC} types, generate complex-valued symbols on the unit circle. While \gls{PS-MP} corresponds to a $2^\mathrm{M}$-\gls{PSK} constellation (where $\mathrm{M}$ defines both sequence length $\mathrm{N}=4^\mathrm{M}$ and constellation size), \gls{ZC} sequences present more challenging optical generation using an electro-optic modulator due to their non-uniform phase distribution.

The remaining sequences (\gls{m-sequence}, Legendre, Golay) employ real-valued \gls{BPSK} modulation over the alphabet $\{-1, +1\}$. For m-sequences and Legendre sequences, a non-binary mapping $\{e^{j\Phi}, +1\}$ with $\Phi = \cos^{-1}\left(-\frac{\mathrm{N}-1}{\mathrm{N}+1}\right)$ preserves perfect autocorrelation properties~\cite{gabidulin_1993_non-binary}. However, as $\Phi \to \pi$ (so $e^{j\Phi} \to -1$) for longer sequences, the simpler real-valued $\{-1, +1\}$ mapping is preferred for practical optical implementation.

All the following tests use a nominal sequence length of approximately $16{.}384$~symbols, with slight variations depending on the sequence type. Specifically, \gls{PS-MP} \gls{CAZAC} and Golay sequences have an exact length of $16{.}384$~symbols. Legendre ($\mathrm{N}=16{.}363$), \gls{ZC} ($\mathrm{N}=16{.}381$), and m-sequence ($\mathrm{N}=16{.}383$) use the closest available sequence lengths satisfying their mathematical constraints.

For dual-polarisation probing, Golay codes employ two mutually orthogonal pairs assigned to the two orthogonal polarisation tributaries. All other sequence types use a cyclic shift of $\mathrm{N}/2$ for the second polarisation tributary. Neglecting noise, the received signals on each polarisation after propagation through a channel with Jones matrix $\mathrm{\mathbf{H}}$ are given by the following discrete-time model:

\begin{equation}
        \begin{aligned}
            &\left\{
            \begin{aligned}
            \mathbf{y_1} &= \mathbf{h_{11}}\ast \mathbf{S_1}+\mathbf{h_{12}}\ast \mathbf{S_2}, \\[4pt]
            \mathbf{y_2} &= \mathbf{h_{21}}\ast \mathbf{S_1}+\mathbf{h_{22}}\ast \mathbf{S_2},
            \end{aligned}
            \right.
        \end{aligned}
    \end{equation}

\noindent where $\ast$ denotes convolution, $\mathbf{y_1}$ and $\mathbf{y_2}$ are the received signals on each polarisation, $\mathbf{h_{ij}}$ are the channel impulse responses from input polarisation $j$ to output polarisation $i$, and $\mathbf{S_1}$ and $\mathbf{S_2}$ are the transmitted sequences (orthogonal pairs for Golay, cyclic shifts otherwise). 

In coded \gls{DAS} systems, channel estimation is performed by correlating the received signal with the originally transmitted code~\cite{dorize_enhancing_2018}. For orthogonal pairs with perfect autocorrelation, channel estimation via correlation yields clean channel recovery, with $\circledast$ denoting the correlation operation:

\begin{equation}
        \begin{aligned}
            &\left\{
            \begin{aligned}
            \mathbf{y_1} \circledast \mathbf{S_1} &= \mathbf{h_{11}}\ast \delta(n), \\[4pt]
            \mathbf{y_1} \circledast \mathbf{S_2} &= \mathbf{h_{12}}\ast \delta(n), \\[4pt]
            \mathbf{y_2} \circledast \mathbf{S_1} &= \mathbf{h_{21}}\ast \delta(n), \\[4pt]
            \mathbf{y_2} \circledast \mathbf{S_2} &= \mathbf{h_{22}}\ast \delta(n).
            \end{aligned}
            \right.
        \end{aligned}
\end{equation}

For sequences where the second polarisation is generated by a cyclic shift, each correlation also produces a channel estimation replica at $\mathrm{N}/2$ offset:

\begin{equation}
        \begin{aligned}
            &\left\{
            \begin{aligned}
            \mathbf{y_1} \circledast \mathbf{S_1} &= \mathbf{h_{11}}\ast \delta(n) + \mathbf{h_{12}}\ast \delta \Big(n-\frac{\mathrm{N}}{2}\Big), \\[4pt]
            \mathbf{y_1} \circledast \mathbf{S_2} &= \mathbf{h_{11}}\ast \delta \Big(n-\frac{\mathrm{N}}{2}\Big) + \mathbf{h_{12}}\ast \delta(n), \\[4pt]
            \mathbf{y_2} \circledast \mathbf{S_1} &= \mathbf{h_{21}}\ast \delta(n) + \mathbf{h_{22}}\ast \delta\Big(n-\frac{\mathrm{N}}{2}\Big), \\[4pt]
            \mathbf{y_2} \circledast \mathbf{S_2} &= \mathbf{h_{21}}\ast \delta\Big(n-\frac{\mathrm{N}}{2}\Big) + \mathbf{h_{22}}\ast \delta(n).
            \end{aligned}
            \right.
        \end{aligned}
\end{equation}

Therefore, sequences with perfect or near-perfect autocorrelation properties—and consequently a high \gls{PSR}—are highly desirable for achieving accurate channel recovery, especially when the estimated channel presents a large dynamic range, which is the case in \gls{DAS} applications due to coherent fading. The following subsections present the correlation functions and time–frequency characteristics of each sequence type, serving as key indicators of their expected performance.

\subsection{Correlation analysis in ideal conditions}\label{sec:seq_b2b_corr}

This section analyses the autocorrelation and cross-correlation characteristics of the previously introduced sequences. Cross-correlation is computed between sequences transmitted over the two orthogonal polarisation tributaries. All (auto)correlation properties refer to periodic (auto)correlation, appropriate for cyclic sequence transmission in DAS systems.

\begin{figure}[!t]
    \centering
    \includegraphics[width=0.48\textwidth]{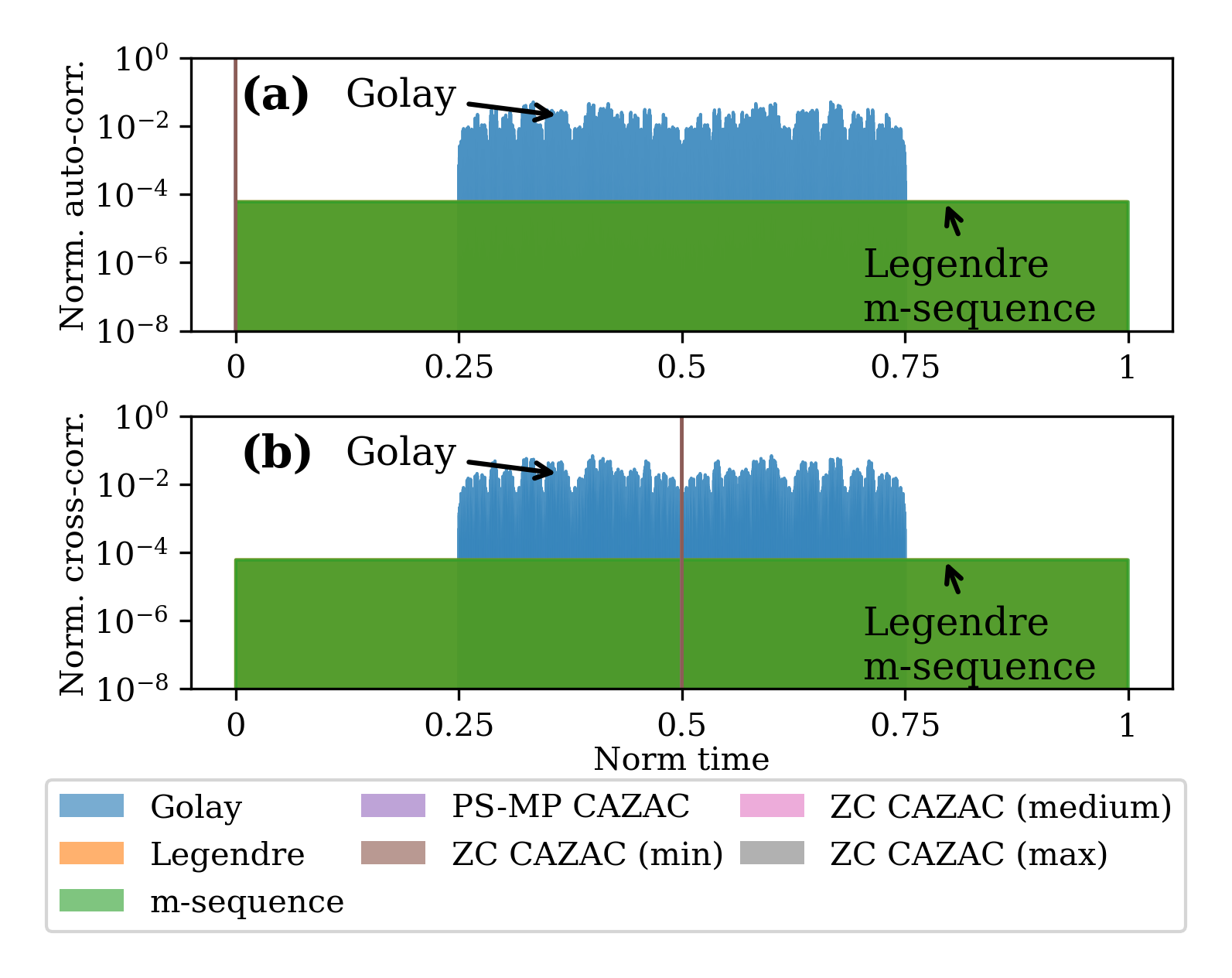}
    \caption{(a) Autocorrelation and (b) cross-correlation for all considered coded sequences.}
    \label{fig:b2b-correlations}
\end{figure}

Figure~\ref{fig:b2b-correlations}a shows the normalised periodic autocorrelation traces versus the normalised time for all sequences. All exhibit a Dirac delta peak at $t=0$, but sidelobe behaviour varies significantly. Legendre and \gls{m-sequence} show near-ideal autocorrelation, though imperfect due to \gls{BPSK} $\{-1,+1\}$ mapping. \gls{PS-MP} and \gls{ZC} \gls{CAZAC} achieve perfect autocorrelation (zero sidelobes). Golay complementary pairs also show perfect autocorrelation within the 0.25 normalised time range, but sidelobes reappear beyond this limit, constraining maximum fibre length compared to flat \gls{CAZAC} response~\cite{dorize_optimal_2020}.

Figure~\ref{fig:b2b-correlations}b illustrates the cross-correlation traces, which exhibit similar properties to the autocorrelation. By construction, the y-polarisation is generated by applying a cyclic shift of $\mathrm{N}/2$ to the x-polarisation sequence, producing a Dirac delta peak at the centre of the cross-correlation for all sequences except those using mutually orthogonal pairs. Golay complementary pairs use mutually orthogonal sequences for each polarisation, eliminating the cross-correlation delta peak.

\subsection{Time-frequency response}\label{sec:seq_b2b_time_freq}

\begin{figure}[!t]
    \centering
    \includegraphics[width=0.48\textwidth]{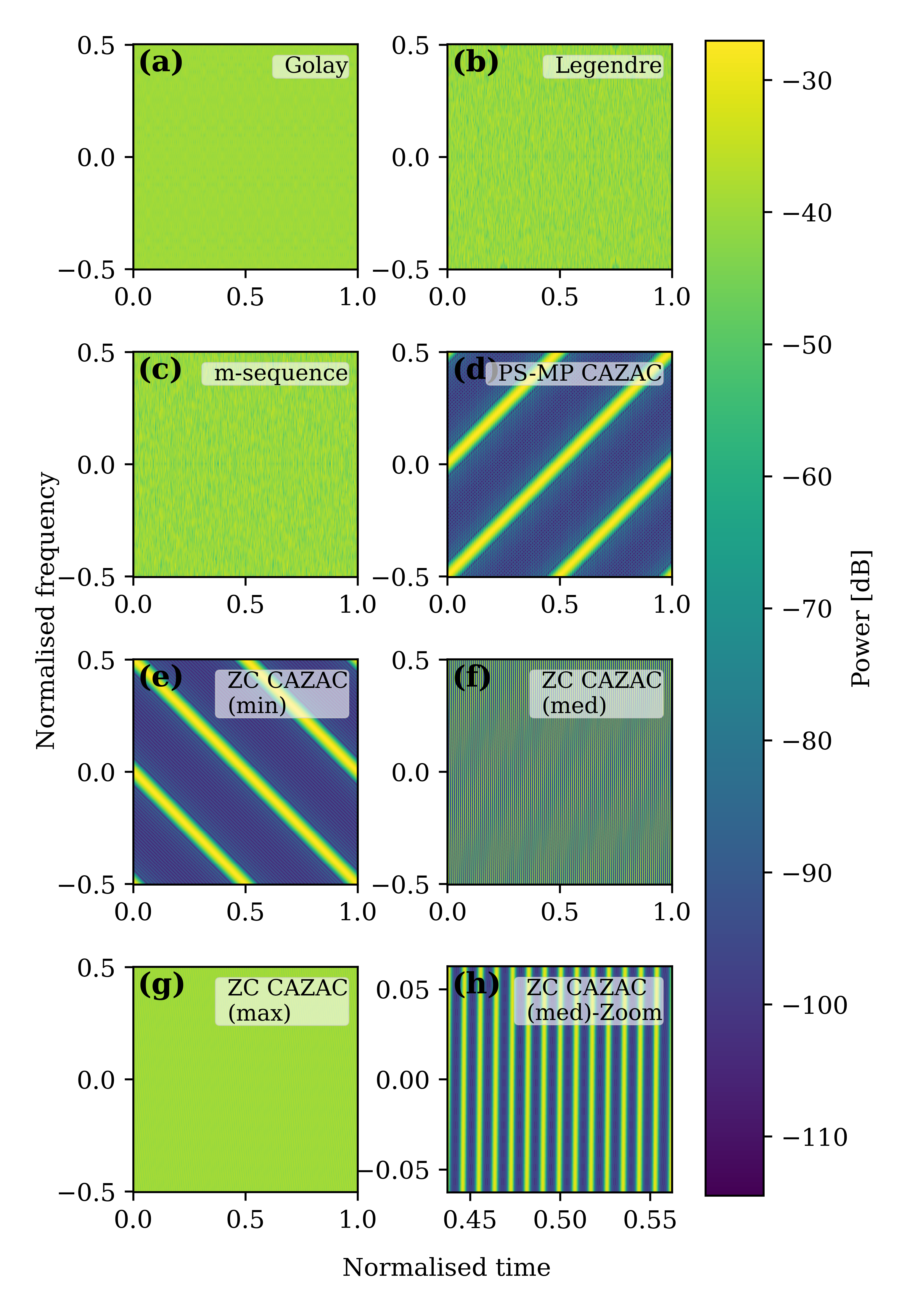}
    \caption{Spectrograms of all considered coded sequences. Subplots (a-g) show full normalised spectrograms; subplot (h) shows the zoomed central region of the \gls{ZC}-\gls{CAZAC} medium frequency diversity case.}
    \label{fig:b2b-t-f}
\end{figure}

Figure~\ref{fig:b2b-t-f} illustrates the short-time Fourier transform spectrograms (64-sample window) for all sequences. The horizontal axis represents normalised time, the vertical axis normalised frequency, and colour intensity indicates instantaneous spectral power.

Frequency diversity measures how broadly a sequence's energy spreads across frequencies (from narrow to widely uniform). Most sequences (Golay, Legendre, \gls{m-sequence}) exhibit high frequency diversity, with energy distributed broadly across all frequencies throughout the time axis, i.e. many spectral components are excited simultaneously. In contrast, \gls{PS-MP} and \gls{ZC} \gls{CAZAC} sequences show frequency-swept behaviour (Fig.~\ref{fig:b2b-t-f}d-e). For \gls{ZC} sequences, frequency diversity is tunable via parameter $q$ in Eq.~\ref{eq:zc}, spanning minimal to maximal frequency diversity (Fig.~\ref{fig:b2b-t-f}e-g; Fig.~\ref{fig:b2b-t-f}h zooms Fig.~\ref{fig:b2b-t-f}f).

\subsection{Correlation analysis in the presence of laser phase noise}\label{sec:seq_b2b_corr_pn}

This section investigates the effect of phase noise introduced by the laser compared with the ideal case in Section~\ref{sec:seq_b2b_corr}. To this end, the correlation is computed between the original undisturbed sequences ($s[n]$) and sequences impaired by phase noise ($s'[n]$):

\begin{equation}\label{eq:pn_affected_sequence}
    s'[n] = s[n] e^{j \phi_{TX}[n]} e^{j \phi_{LO}[n]},
\end{equation}

\noindent where $\phi_{TX}[n]$ and $\phi_{LO}[n]$ denote the transmitter and \gls{LO} phase noise, respectively. Since the same laser is used for both, the transmitter phase noise is modelled as a delayed version of the \gls{LO} phase noise, which follows a random walk process.

Although these simulations correspond to a back-to-back scenario without fibre propagation, the effect of phase noise can still be related to the round-trip delay corresponding to different fibre segments. Shorter delays represent segments closer to the receiver, while longer delays correspond to more distant points along the fibre. To emulate this relationship, two transmitter–\gls{LO} delays are considered: 1 and 1000~symbols—both much shorter than the sequence length, defined in Section~\ref{sec:selected_sequences}.

\begin{figure}[!t]
    \centering
    \includegraphics[width=0.48\textwidth]{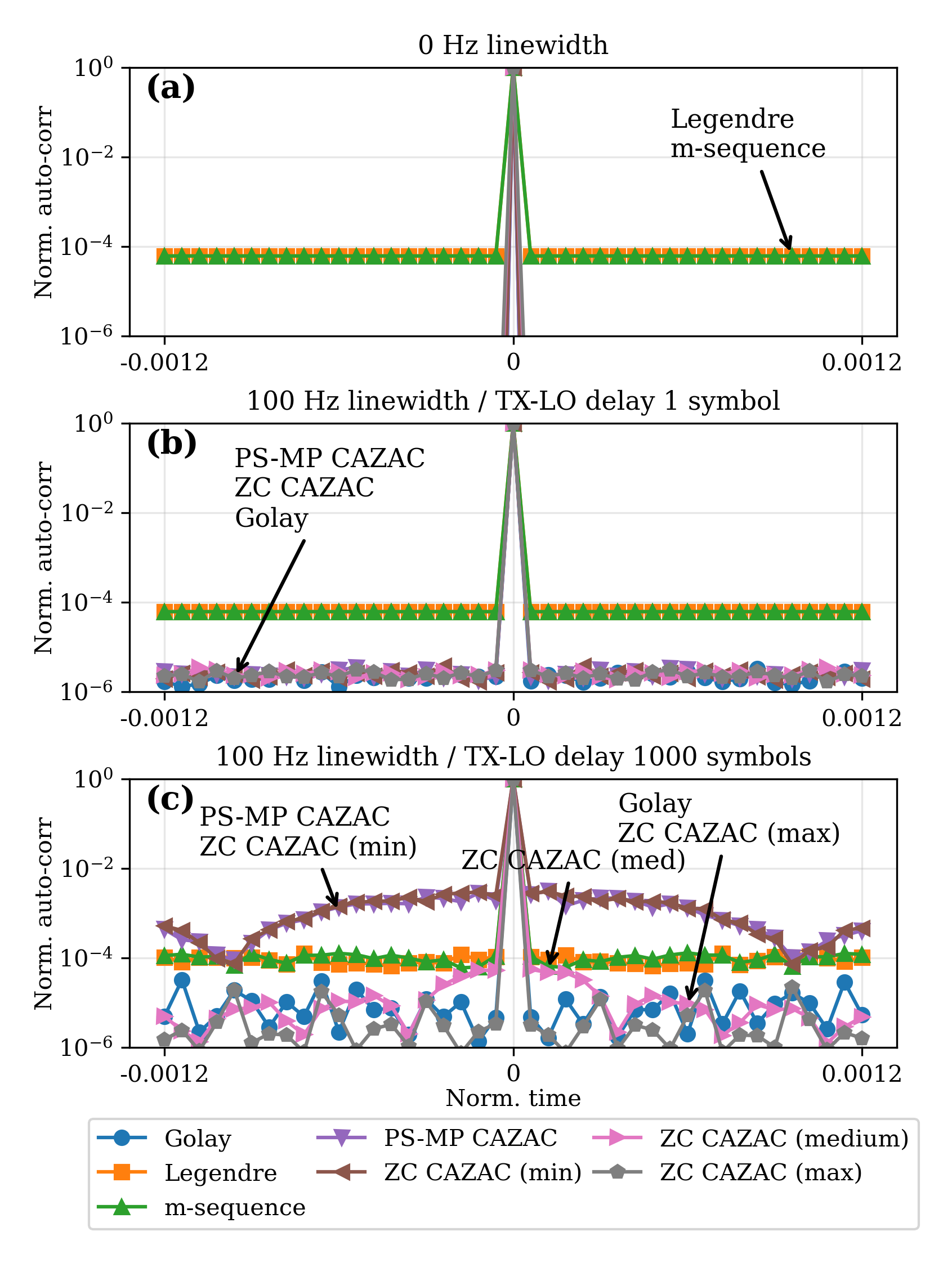}
    \caption{Correlation of each of the original considered coded sequences with the phase noise-affected sequence: (a) 0~Hz linewidth, (b) 100~Hz linewidth and 1~symbol TX-LO delay and (c) 100~Hz linewidth and 1000~symbols TX-LO delay.}
    \label{fig:b2b-pn-correlation}
\end{figure}

Figure~\ref{fig:b2b-pn-correlation} presents the normalised correlation results. First, Fig.~\ref{fig:b2b-pn-correlation}a shows the ideal case with \qty{0}{\hertz} laser linewidth (no phase noise), serving as a baseline. This figure is a zoom around the Dirac delta peak of Fig.~\ref{fig:b2b-correlations}a. Only Legendre and \gls{m-sequence} exhibit visible sidelobes due to their imperfect autocorrelation.

Figures~\ref{fig:b2b-pn-correlation}b-c show the same traces for \qty{100}{\hertz} laser linewidth, for two different transmitter-\gls{LO} delays. Figure~\ref{fig:b2b-pn-correlation}b corresponds to a 1-symbol transmitter-\gls{LO} delay. Legendre, and \gls{m-sequence} maintain their sidelobe levels, while \gls{PS-MP} \gls{CAZAC}, \gls{ZC} \gls{CAZAC}, and Golay lose perfect autocorrelation, developing sidelobes above $10^{-6}$.

For the 1000~symbol delay in Fig.~\ref{fig:b2b-pn-correlation}c, degradation becomes more pronounced. Legendre and \gls{m-sequence} remain largely unaffected, but \gls{PS-MP} \gls{CAZAC} and \gls{ZC} \gls{CAZAC} minimum frequency diversity now exhibit the highest sidelobes, consistent with their frequency-sweeping spectrograms (Fig.~\ref{fig:b2b-t-f}d-e). Golay complementary sequences and \gls{ZC} \gls{CAZAC} maximum frequency diversity show the lowest sidelobes, with \gls{ZC} \gls{CAZAC} medium frequency diversity performing intermediately. These results suggest that higher frequency diversity provides robustness against phase noise degradation.

\section{Sequence comparison on a DAS system}\label{sec:seq_das}

\begin{figure}[!t]
	\centering	
	\includegraphics[width=0.45\textwidth]{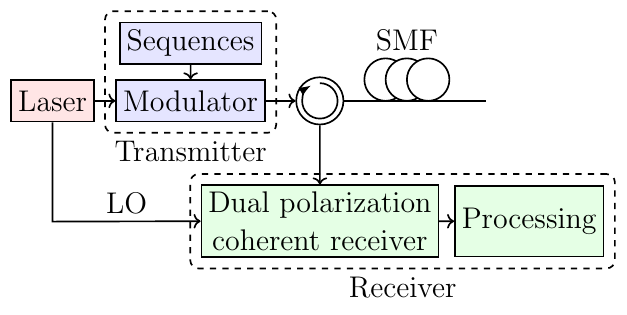}%
	\caption{Block diagram of the simulation framework.}\label{fig:das-setup}
\end{figure}

Figure~\ref{fig:das-setup} illustrates the simulation framework used to evaluate the performance of the different coded sequences. The framework provides a common basis for analysing the system response under various signal designs and laser phase noise conditions. In this configuration, continuous-wave light is modulated with coded sequences, \gls{BPSK} for $\{-1, +1\}$ sequences or complex symbols directly for \gls{CAZAC} sequences. Dual-polarisation probing employs orthogonal sequences across both polarisations. The only transmitter impairment considered is laser phase noise, which is modelled as a Wiener process~\cite{magarini_empirical_2011}.

Signal propagation along the \gls{SMF} channel is modelled using the Rayleigh backscattering model, incorporating fibre attenuation but neglecting \gls{RSOP}~\cite{liokumovich2015fundamentals, damask2004construction}. At the coherent receiver, sharing the transmitter laser, additional \gls{LO} phase noise and \gls{AWGN} with zero-mean and variance $\sigma_n^2 = 4.14 \times 10^{-9}$ per quadrature are introduced. \gls{LO} phase noise ($\phi_{LO}[n]$) is modelled as a time-advanced version of the transmitter laser Wiener process ($\phi_{TX}[n]$):

\begin{equation}\label{eq:phi_tx_lo_def}
    \begin{split}
    \phi_{TX}[n] &= \phi[n-\mathrm{D}],\\
    \phi_{LO}[n] &= -\phi[n],
    \end{split}
\end{equation}

\noindent where $\phi[n]$ represents the common laser phase noise, modelled as a random walk process, and $\mathrm{D}$ denotes the delay between transmission and reception. For the tests presented here, this delay is fixed at approximately \qty{20}{\nano\second}, corresponding to a \qty{4}{\meter} path length difference in \gls{SMF}. The negative sign in $\phi_{LO}[n]$ arises from the coherent detection process, in which the received optical field is mixed with the \gls{LO} field, resulting in a conjugate phase term for the \gls{LO} contribution.

Post-processing correlates received signals with transmitted sequences to estimate the $2 \times 2$ channel matrix ($\mathbf{\hat{H}}$). The phase is computed as  $0.5 \angle \det(\mathbf{\hat{H}})$, and phase differentials are derived between consecutive fibre segments to monitor any events~\cite{dorize_enhancing_2018}. In the simulations presented here, a symbol rate of \qty{50}{\mega\baud} and the same sequence length as in the previous sections (approximately $16.384$~symbols) are used, resulting in approximately \qty{2}{\meter} per fibre segment.

For the comparison of sequence performance, two evaluation metrics are considered:

\begin{enumerate}
	\item Relative determinant error:
    \begin{equation}
        \frac{|\det(\hat{\mathbf{H}}) - \det(\mathbf{H})|}{\det(\mathbf{H})}.
    \end{equation}
    
	\item Phase error: 
    \begin{equation}
        |\angle \det(\mathbf{\hat{H}}) - \angle \det( \mathbf{H})|,
    \end{equation}
    
\end{enumerate}

\noindent where $\mathbf{H}$ denotes the true fibre channel matrix used in the simulation.

The following sections analyse the received signal intensities and assess the performance of the different coded sequences under various laser phase noise conditions.

\subsection{Received intensities}

\begin{figure}[!t]
	\centering	
	\includegraphics[width=0.485\textwidth]{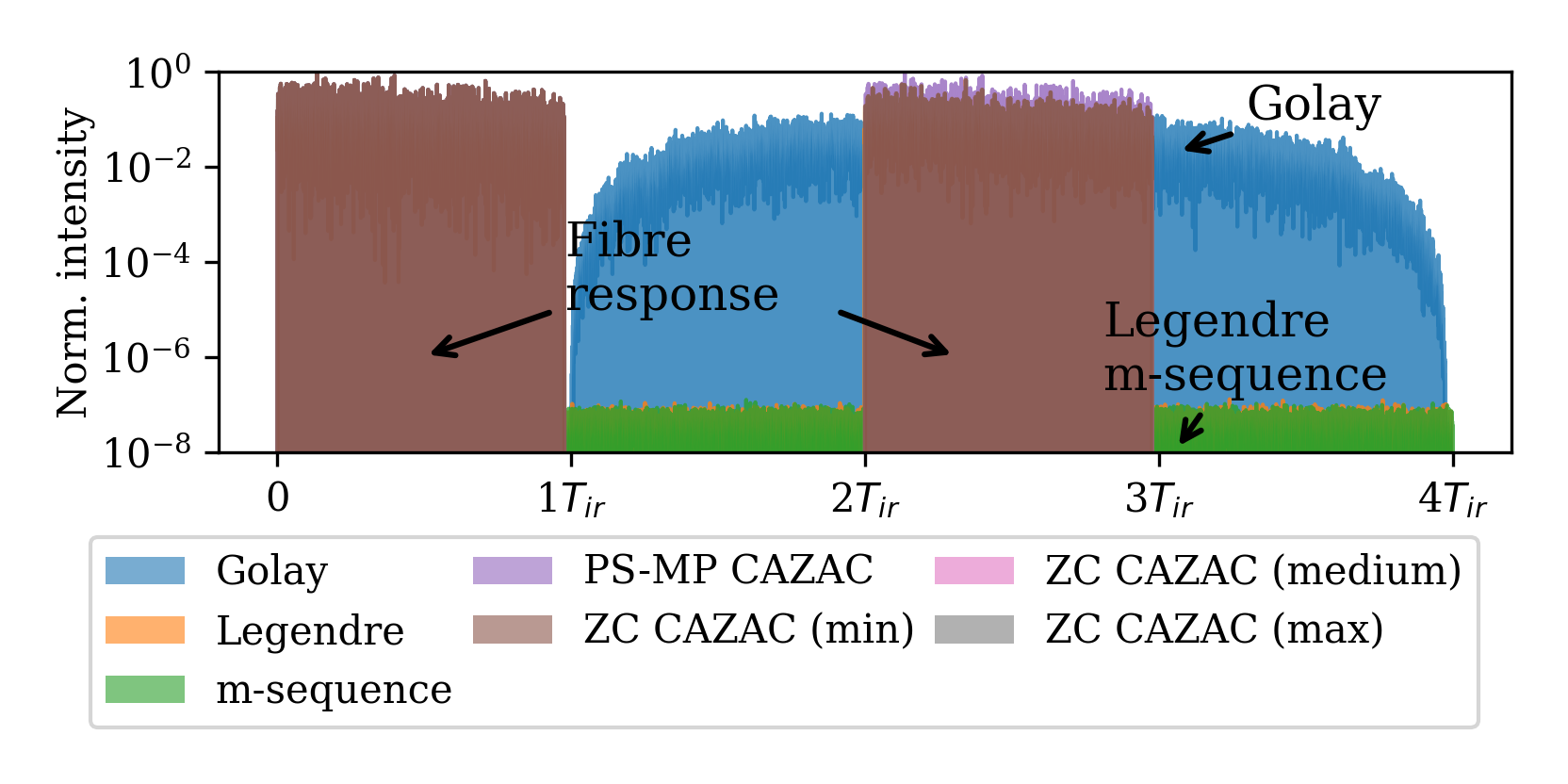}%
	\caption{Normalised received intensities for all considered coded sequences.}\label{fig:das-det}
\end{figure}

Figure~\ref{fig:das-det} presents the normalised received intensity  for all considered sequences coming from a \qty{8.4}{\kilo\meter} fibre, in the case where no laser phase noise is introduced and only \gls{AWGN} at the receiver side is considered. Compared with the results in Fig.~\ref{fig:b2b-correlations}, where a delta-like correlation peak was observed, the figure now shows the complete fibre impulse response.

For $\mathrm{N}/2$-cyclically-shifted probing sequences, accurate estimation of the full Jones matrix for round-trip propagation requires the sequence duration ($\mathrm{T_{code}}$) to exceed twice the fibre channel spreading time ($\mathrm{T_{ir}}$), that is, $\mathrm{T_{code}} > 2 \mathrm{T_{ir}}$ to avoid the overlap between the consecutive estimates of the channel. This condition ensures that the entire backscattered response fits within one interrogation period, as illustrated in Fig.~\ref{fig:das-det}. For Golay sequences, an additional factor of two is required ($\mathrm{T_{code}} > 4 \mathrm{T_{ir}}$) to avoid aliasing effects caused by their non-zero autocorrelation zone~\cite{dorize_enhancing_2018}.

For \gls{CAZAC}, Legendre and \gls{m-sequence}, orthogonal sequences for dual-polarisation probing are generated by applying a cyclic shift. This approach produces a secondary replica of the channel response at $2\mathrm{T_{ir}}$. 

Out-of-band intensities match the patterns in the ideal correlations from Fig.~\ref{fig:b2b-correlations}a-b: \gls{CAZAC} and Golay sequences show imperceptible sidelobes, while Legendre and \gls{m-sequence} exhibit visible noise floors due to their near-ideal but imperfect autocorrelation.

\subsection{Performance under different laser phase noise conditions}\label{sec:seq_das_noise}

Simulations are performed for fibre lengths ranging from 1 to \qty{20}{\kilo\metre}, extending beyond the theoretical maximum reach of the different code types. To assess the performance of each code, the relative determinant error and the phase error with respect to the true simulated channel are evaluated. These errors are computed only on a subset of segments: one out of every ten consecutive fibre segments is selected based on reflectivity, yielding an effective gauge length of approximately \qty{20}{\meter}.

For each probing method and fibre length, five simulations are carried out using different fibre realisations, i.e. distinct Rayleigh backscattering profiles with varying scatterer magnitudes and phases. The resulting relative determinant and phase errors are then averaged over all selected fibre segments and all realisations.

When only \gls{AWGN} is present, the \gls{SNR} estimated after correlation ranges between \qty{30}{}-\qty{39}{\decibel} for a \qty{500}{\meter} fibre, depending on the correlation properties of the probing sequence. Sequences with imperfect correlation properties yield \qty{30}{\decibel}, while perfect correlation sequences reach \qty{39}{\decibel}.

The estimated \gls{SNR} is calculated using $|\det{\mathbf{\hat{H}}}|$ as an estimator of Rayleigh backscattered intensity, valid when $\hat{\mathbf{H}}$ can be expressed as a common attenuation and phase term multiplying a unitary round-trip matrix~\cite{dorize_2021_identification}:

\begin{equation}
    \mathrm{SNR} = 10 \log_{10} \Bigg(\!\!\frac{\mathbb{E}\{|\det(\mathbf{\hat{H}})|\}}{\mathbb{E}\{|\mathbf{n_{ij}}|^2\}} \!\!\Bigg)\!,
\end{equation}

\noindent where $\mathbf{n_{ij}}$ denotes the noise samples corresponding to each element of the estimated channel matrix  $\hat{\mathbf{H}}$. Each $\mathbf{n_{ij}}$ represents the additive noise contribution of the $(i, j)$ path, measured in a window of correlation outside the channel response. 

\begin{figure}[!t]
	\centering	
	\includegraphics[width=0.48\textwidth]{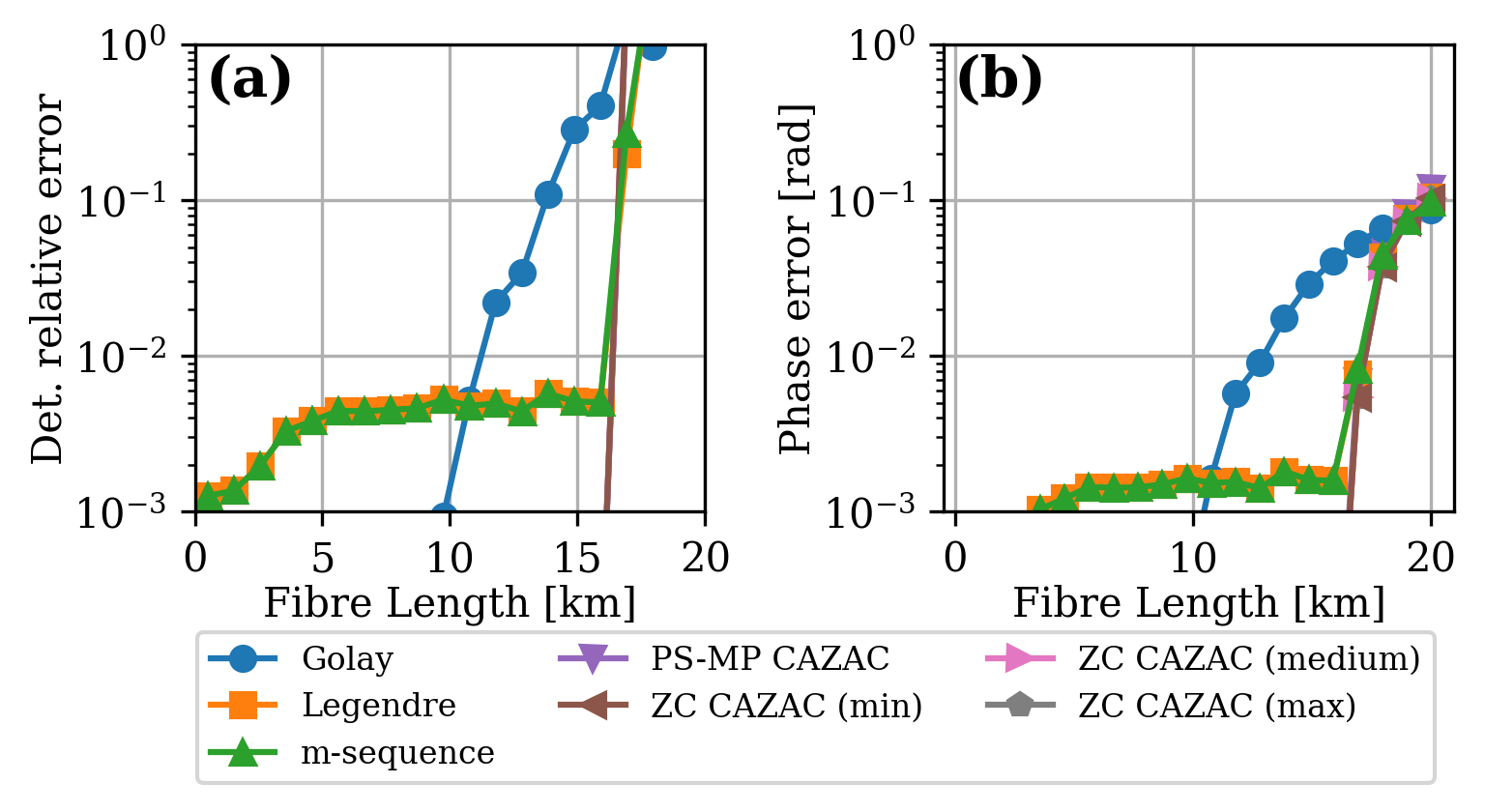}%
	\caption{(a) Determinant relative error and (b) phase error for a scenario with AWGN.}\label{fig:das-det-and-phase-error-awgn}
\end{figure}

Figure~\ref{fig:das-det-and-phase-error-awgn} reports performance metrics for this case, where laser phase noise is not present and only \gls{AWGN} is included. In these conditions, different coded sequences already exhibit clear performance differences. Legendre and \gls{m-sequence} codes underperform relative to Golay and \gls{CAZAC} codes. Golay codes achieve nearly perfect channel estimation for shorter fibre spans but deteriorate sharply once the fibre length exceeds their maximum alias-free range (\qty{8.4}{\kilo\metre}), due to the presence of a non-zero autocorrelation zone. In this region, Golay performance is worse than that of Legendre and \gls{m-sequence}. In contrast, \gls{PS-MP} and \gls{ZC} \gls{CAZAC} sequences maintain near-ideal channel estimation up to \qty{16.8}{\kilo\metre}.

\begin{figure}[!t]
	\centering	
	\includegraphics[width=0.48\textwidth]{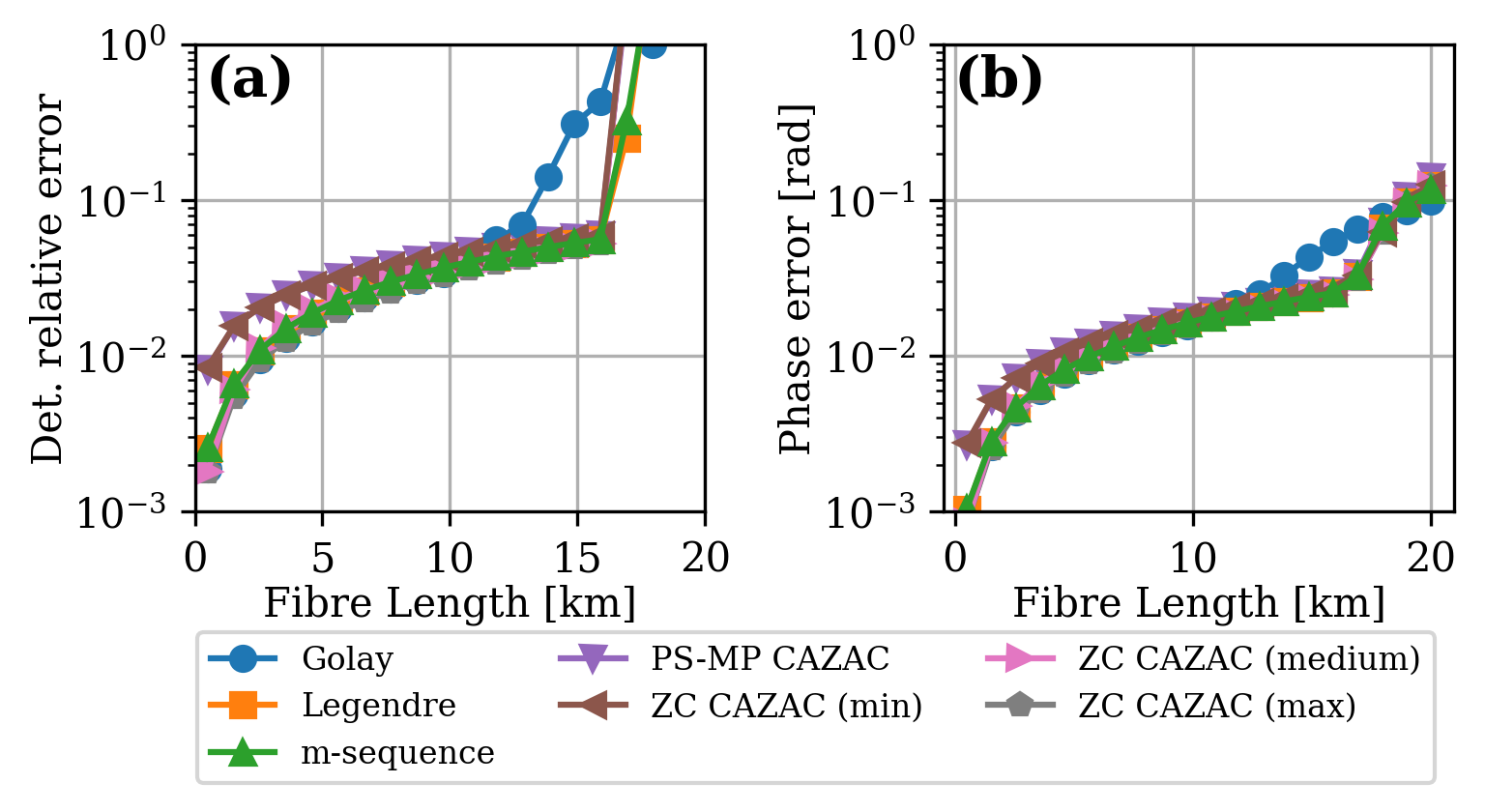}%
	\caption{(a) Determinant relative error and (b) phase error for a scenario with AWGN and phase noise from a 10~Hz laser linewidth.}\label{fig:das-det-and-phase-error-10Hz-pn}
\end{figure}

When laser phase noise is introduced with a laser linewidth of \qty{10}{\hertz}, all sequences exhibit noticeable performance degradation, as shown in Fig.~\ref{fig:das-det-and-phase-error-10Hz-pn}. This degradation is reflected in both increased relative determinant and phase errors. Under these conditions, the performance differences among the various codes become less pronounced. For shorter fibre lengths, \gls{PS-MP} and \gls{ZC} \gls{CAZAC} sequence with minimum frequency diversity show higher errors compared with the other codes. This behaviour is directly linked to their lower frequency diversity. Consequently, Golay, Legendre, \gls{m-sequence}, and the \gls{ZC} codes with higher frequency diversity achieve better overall performance. At \qty{1.5}{\kilo\meter}, high-diversity sequences achieve $2.7\times10^{-3}$ rad phase error compared to $5.3\times10^{-3}$ rad for low-diversity codes. At longer fibre spans, the performance of the Golay sequence begins to degrade due to aliasing effects associated with its finite non-zero autocorrelation zone, while the other sequences, such as Legendre, continue to maintain lower error levels under the same conditions.

\begin{figure}[!t]c
	\centering	
	\includegraphics[width=0.48\textwidth]{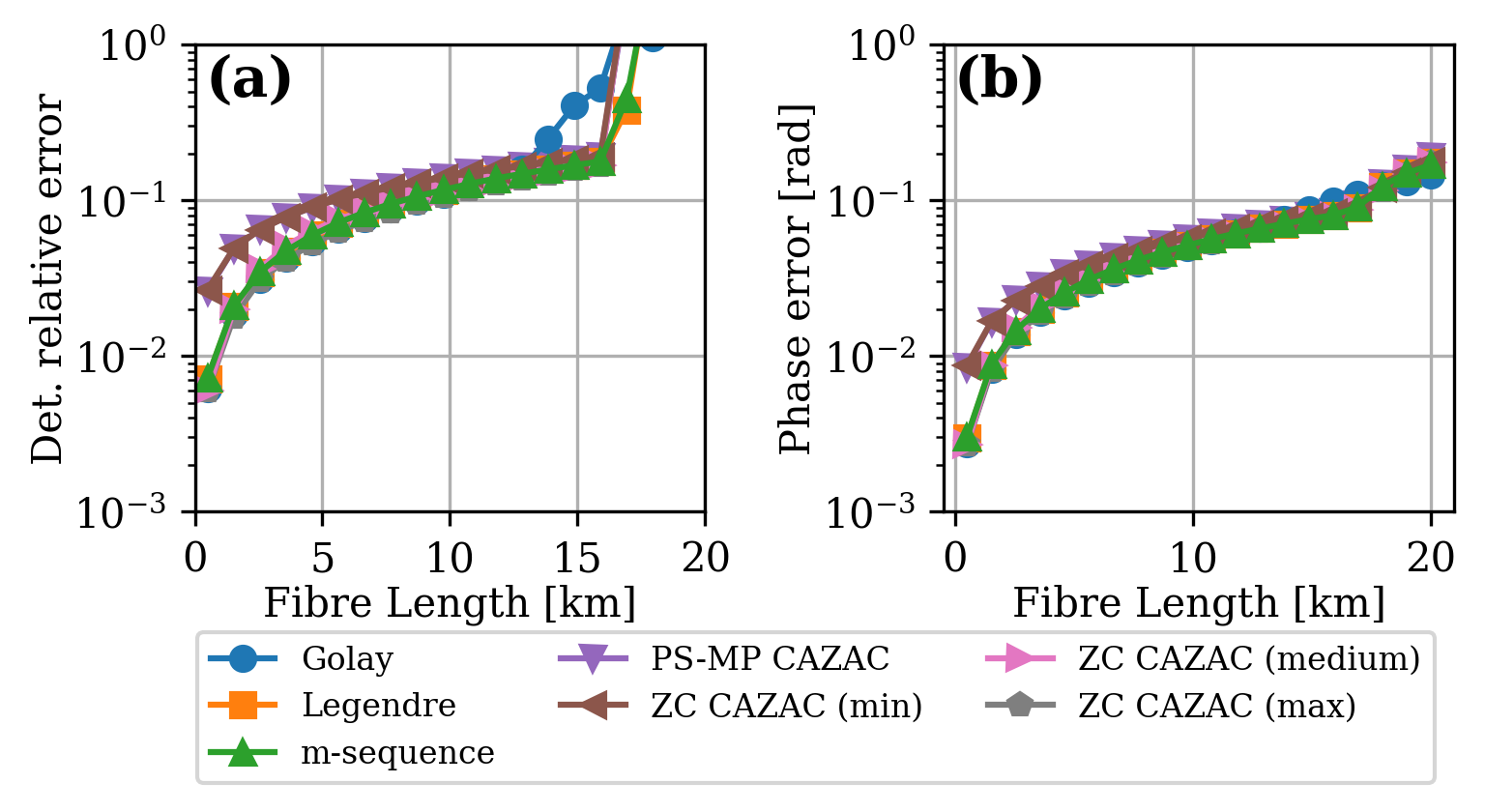}%
	\caption{(a) Determinant relative error and (b) phase error for a scenario with AWGN and phase noise from a 100~Hz laser linewidth.}\label{fig:das-det-and-phase-error-100Hz-pn}
\end{figure}

Further increasing the laser linewidth to \qty{100}{\hertz} (Fig.~\ref{fig:das-det-and-phase-error-100Hz-pn}) results in a more pronounced overall degradation. The performance differences among the various codes become less distinct, while the frequency diversity advantage persists. For the same \qty{1.5}{\kilo\meter} fibre length, we obtain now a $8.5\times10^{-3}$ rad phase error for high frequency diversity compared to $1.7\times10^{-2}$ rad phase error for low frequency diversity sequences.

\section{Conclusions}

We presented a comprehensive analysis of different coded sequences for \gls{DAS} systems to assess their performance under various laser phase noise conditions. 

The study began by characterising the correlation properties, frequency diversity, and phase noise sensitivity of several sequence families, including \gls{CAZAC}, Golay, Legendre and  \gls{m-sequence}. Correlation analysis showed that \gls{CAZAC} sequences exhibit ideal autocorrelation properties, enabling accurate channel estimation over extended fibre lengths. Golay sequences achieved similarly strong results but were limited by their non-zero autocorrelation zone. In contrast, Legendre and \gls{m-sequence} displayed lower correlation performance. Simulation results revealed that the effects of laser linewidth vary across sequence types. While all codes showed autocorrelation degradation with increased linewidth, those with higher frequency diversity maintained lower sidelobes compared to sequences with lower frequency diversity.

This trend was confirmed in the \gls{DAS} system numerical simulations. Under \gls{AWGN} alone, \gls{CAZAC} and Golay sequences maintained near-ideal channel estimation, outperforming Legendre and \gls{m-sequence} codes. With \qty{10}{\hertz} laser phase noise, all sequences degraded, but frequency diversity emerged as the key differentiator: sequences with low frequency diversity showed higher errors at short fibre lengths, while high frequency diversity sequences performed better.

Overall, the results highlight the importance of considering autocorrelation quality and frequency diversity in the design of coded \gls{DAS} for \gls{O-ISAC} applications requiring reliable sensing. With laser phase noise being a limiting performance factor at higher linewidths, future work should also explore phase noise compensation techniques to further extend reliability and reach.

\section{Acknowledgments}
\noindent  This work has received funding from IMT within the FRAME-XG program.

\bibliographystyle{IEEEtran}
\bibliography{references}

\end{document}